\documentclass{article}
\usepackage[final]{colm2026_conference}

\usepackage{microtype}
\usepackage{hyperref}
\usepackage{url}
\usepackage{booktabs}
\usepackage{amsmath}
\usepackage{amssymb}
\usepackage{xspace}
\usepackage{multirow}

\definecolor{darkblue}{rgb}{0, 0, 0.5}
\hypersetup{
  colorlinks=true,
  citecolor=darkblue,
  linkcolor=darkblue,
  urlcolor=darkblue,
  pdftitle={Cross-Benchmark Generalization in Long-Horizon Agents},
  pdfauthor={Sushant Mehta, Logan Ritchie, Liudas Panavas, and Edwin Chen}
}

\newcommand{\env}{LHMTA\xspace}
\newcommand{\tnb}{$\tau^2$-Bench\xspace}
\newcommand{\base}{Qwen3.5-122B-A10B\xspace}
\newcommand{\pp}{\,pp\xspace}
\newcommand{\passk}[1]{pass@#1}

\title{Cross-Benchmark Generalization in Long-Horizon Agents}

\author{%
\begin{tabular}{@{}l@{}}
\bfseries Sushant Mehta\thanks{Corresponding author: \texttt{sushantmehta@surgehq.ai}.}
\quad Logan Ritchie
\quad Liudas Panavas
\quad Edwin Chen\\
{\normalfont Surge AI}
\end{tabular}%
}

\begin{document}

\maketitle
\lhead{Accepted at the COLM 2026 Workshop on Agent Behavior}

\begin{abstract}
For reinforcement learning (RL) in self-contained environments, a policy can get rewards by exploiting environment-specific regularities (tool schemas, grader parsing, task templates) rather than by acquiring transferable skill, and an in-distribution holdout shares those regularities. We argue that the discriminating question is behavioral, namely how a trained agent acts, and that cross-benchmark transfer is the right place to look for it. We post-train an open-weight mixture-of-experts model (\base) on 363 long-horizon Model Context Protocol (MCP) tasks across 27 categories, using a two-stage SFT-then-RL pipeline. Toolathlon performance informed the initial base-family and SFT-teacher choices, but no external-benchmark task or grader entered training and no external score informed the reward, training hyperparameters, trained-checkpoint selection, or stopping. At greedy \passk{1}, the trained model improves over the base on five reported external evaluations: Toolathlon ($+9.6$\pp), \tnb ($+5.3$\pp), BFCL-V4 ($+3.5$\pp), SWE-Bench Pro ($+5.8$\pp), and Terminal-Bench~2 ($+2.8$\pp). Both software-engineering benchmarks improve despite the training collection containing no software-engineering tasks. An exploratory paired-trajectory analysis identifies four recurring behavioral differences (more careful local-goal formation, building goal-relevant working state, keeping parent goals stable through local repairs, and verifying completion) that appear in analogous forms across office workflows and code. These results provide descriptive evidence that long-horizon multi-tool post-training can change ways of working that transfer beyond its training domain.
\end{abstract}

\section{Introduction}

The dominant way to report progress on AI agents is capability-centric: an agent is trained, its task-success rate is measured, and a higher number is taken as evidence of a better agent. For reinforcement learning (RL) on a curated, self-built environment, the success rate alone is a weak signal of what was learned. A policy optimized against a fixed environment may exploit consistent features of that environment's reward landscape (idiosyncrasies of the tool surface, grader implementation, or task templates) rather than improve the underlying skills the environment is meant to teach. Crucially, the natural internal control, a held-out split of the same environment, does not exclude this possibility: an in-distribution holdout shares many environment-specific degrees of freedom with the training set.

This is fundamentally a question about agent behavior: not only what an agent achieves but how it achieves it, and whether the ``how'' is a reusable way of working or a brittle exploit. We take two complementary positions. First, transfer should be tested on external task instances and graders, with results withheld from reward design, training hyperparameters, trained-checkpoint selection, and stopping. Second, the interpretation should be behavioral: we should point to recurring changes in how the agent acts, ask whether those changes appear across superficially different domains, and identify where they do not.

We study both. We post-train an open-weight mixture-of-experts (MoE) model, \base, on Long-Horizon Multi-Tool Agent tasks (\env): 363 training tasks spanning 27 categories of realistic professional workflows, exposed through Model Context Protocol (MCP) servers \citep{mcp2024} and scored by deterministic per-criterion graders. We then evaluate on five external benchmarks whose task instances and graders were excluded from training and analyze how the observed behavior changed.

Our contributions are:
\begin{itemize}
  \item \textbf{An evaluation stance and a consistent transfer result.} We treat the in-distribution holdout as a pipeline check and report five externally maintained benchmark evaluations. The trained model improves on Toolathlon, \tnb, BFCL-V4, SWE-Bench Pro, and Terminal-Bench~2, including two software-engineering benchmarks despite no software-engineering training tasks (Section~\ref{sec:results}).
  \item \textbf{A behavioral account of the transfer.} Through paired base-vs-trained trajectory analysis we identify four recurring behavioral differences (goal formation, goal-relevant state-building, goal stability, and verification), show that each appears across general tool use and code, and quantify related signals on software trajectories (Section~\ref{sec:behavior}).
  \item \textbf{Calibrated reporting.} We disclose the complete reward, external-evaluation firewall, diagnostic-result provenance, exploratory nature of the trajectory analysis, and single-run limitation (Sections~\ref{sec:method}--\ref{sec:behavior}).
\end{itemize}

\section{Related Work}

\paragraph{RL for agents and verifiable rewards.} RL with verifiable rewards has become the standard recipe for eliciting multi-step reasoning and tool use from language models \citep{deepseekr1_2025,deepseekmath2024}. Group-relative estimators such as GRPO \citep{deepseekmath2024} and the sequence-level GSPO \citep{gspo2025} we use avoid a learned value model by normalizing rewards within a group of rollouts. Most open work in this line trains on a single domain (typically code or math) and reports in-domain gains; our focus is whether the behaviors learned in one environment transfer outside it.

\paragraph{Coding-agent RL and environments.} A productive recent line grounds code LLMs in execution feedback \citep{rlef2024,swerl2025} and builds executable training environments and verifiers for software agents \citep{swegym2024,r2egym2025,skyrl2025}. These environments are powerful but largely code-centric. \env\ instead spans general professional software work (office suites, terminals, calendars, file systems, web), and we ask what transfers from that broader distribution to SWE-Bench Pro \citep{swebenchpro2025}, a repository-level issue-resolution benchmark in the SWE-bench lineage \citep{swebench2024}, and Terminal-Bench~2 \citep{terminalbench2026}, which evaluates difficult work in command-line environments.

\paragraph{Agentic benchmarks.} Realistic, long-horizon agent benchmarks include web environments \citep{webarena2024}, broad agent suites \citep{agentbench2024}, tool-agent-user interaction benchmarks \citep{taubench2024,tau2bench2025}, function-calling evaluation \citep{bfcl2025}, and long-horizon tool execution \citep{toolathlon2025}. We use external tasks and graders from \tnb, BFCL-V4, and Toolathlon to test whether changes extend beyond \env; Section~\ref{sec:protocol} states the known Toolathlon surface overlap.

\paragraph{Cross-domain post-training.} Recent work directly tests whether agentic post-training gains extend beyond the training distribution, including high-fidelity enterprise environments, unseen-domain reinforcement post-training, agent reinforcement fine-tuning, and tool-integrated RL \citep{corecraft2026,hu2026breaking,xi2026agentgeneralization,chen2025toolgeneralization}. Our focus is complementary: we connect cross-benchmark score changes to an exploratory account of recurring trajectory-level behavior.

\paragraph{Data selection and what makes data generalize.} Instruction-tuning research moved from ``more data'' to ``better-selected data'': selecting prompt-sensitive tasks \citep{activeit2023}, balancing complexity/quality/diversity \citep{deita2024}, and choosing examples by their influence on a target capability \citep{less2024}. A recurring finding is that surface difficulty and surface diversity are poor proxies; what matters is the underlying reasoning a task exercises. Our behavioral analysis is the agentic analogue of that lesson: we characterize the task pressures an environment exerts rather than its surface domains.

\paragraph{Behavior over scores.} Our framing aligns with the behavioral study of machines \citep{rahwan2019machine}: we seek trajectory-level behavioral explanations of how an agent acts, not only aggregate success. It also builds on failure-derived accounts of agentic capabilities \citep{ritchie2026hierarchy}, process-supervision work \citep{lightman2024}, and parameter-efficient studies of what fine-tuning does and does not change \citep{lora2022,loraless2024}.

\section{The Training Environment}
\label{sec:env}

\env\ is a suite of long-horizon agentic tasks designed to exercise realistic professional work rather than single tool calls. It spans 27 categories covering office suites (spreadsheets, documents, slides), terminals and \texttt{bash}, calendars and scheduling, search and retrieval, file systems, and browser automation; it contains no software-engineering tasks. Each environment is exposed as one or more MCP servers \citep{mcp2024}, so a single trajectory routinely composes several servers (for example, a browser, a spreadsheet reader, a file system, and a calendar) inside one rollout.

\paragraph{Verifiable, per-criterion graders.} Every task ships a deterministic Python grader that scores completion from the \emph{final} environment state (e.g., the contents of files the agent was asked to produce). Graders are structured per criterion (e.g., 8 of 10 sub-conditions satisfied). This structure is central to our reward design (Section~\ref{sec:method}): it supports both a sparse pass/fail reward and a dense partial-completion reward, and it raises the fraction of tasks that yield usable training signal.

\paragraph{Long-horizon and multi-tool.} Successful trajectories typically take 30--40 tool-calling turns and 80K--100K tokens, stressing planning and context management rather than one-shot tool selection. This length also drives real training cost: full trajectories above 50K tokens substantially increase context-parallel sharding overhead, making the RL runs compute-intensive.

\paragraph{Task construction.} Tasks were authored to produce meaningful agentic failures rather than trivial format errors: creators targeted work that strong models could not reliably complete, and the collection was monitored for diversity of tools, task types, and failure modes. Of an initial 430 environment/task pairs, we excluded 27 that depended on rate-limited external APIs (e.g., mapping, calendar, and finance services) or that failed to run deterministically. This left 403 usable tasks: 363 for training and a disjoint 40-task in-distribution holdout. We discuss the implications of this curation-by-failure-mode philosophy, and its limits, in Section~\ref{sec:behavior}.

\section{Training Method}
\label{sec:method}

\paragraph{Base model and adapter.} We post-train \base \citep{qwen35_122b_2026}, an open-weight MoE checkpoint with 122B total and roughly 10B active parameters, chosen as a favorable size/quality trade-off among open models at project start.\footnote{We use the released checkpoint without modification and refer to it by its public name; the contribution here is the environment and the analysis, not the base model.} We use LoRA \citep{lora2022} (rank 32, $\alpha=32$) on all attention and MLP projections (\texttt{q,k,v,o,gate,up,down}). LoRA is motivated both by compute efficiency and by evidence that low-rank adaptation forgets less of the base model's capabilities than full fine-tuning \citep{loraless2024}.

\paragraph{A two-stage recipe (SFT then RL).} The base policy rarely solved these tasks even across four attempts, so binary rewards were too sparse to support RL from that initialization. We therefore used a supervised warm-up before RL. The open-weight teacher, Kimi K2.6 \citep{kimi26_2026}, generated trajectories only on the 363 \env\ training tasks; we retained 3{,}000 trajectories scoring above $0.9$. We then ran RL on the same 363 tasks with the GSPO sequence-level estimator \citep{gspo2025}. We evaluate the one-epoch output of the complete SFT+RL pipeline, called the \emph{trained model} throughout; exploratory continuation through later epochs was not used for the reported evaluations. This experiment was not designed to attribute the observed transfer to either stage in isolation.

\paragraph{Dense criterion reward and low-score effort floor.} Let $r_g\in[0,1]$ be the fraction of deterministic grader criteria satisfied. For trajectories with $r_g<0.5$, the implemented reward also had an effort floor
\[
e=\min(0.30,0.05\min(n_{\mathrm{calls}},6))
  +0.05\,\mathbb{1}_{\mathrm{normal}}
  +0.05\,\mathbb{1}_{\geq 3\ \mathrm{distinct\ tools}},
\qquad
r=\max(r_g,e);
\]
for $r_g\geq0.5$, $r=r_g$. Thus the floor was at most $0.40$ and was a lower bound, not a bonus added to successful grader reward. In a four-attempt diagnostic, $16.8\%$ of tasks had at least one full pass, whereas $82.7\%$ yielded at least one nonzero dense or effort-shaped reward; mean reward increased from $0.30$ to $0.51$. Reward was assigned once per full trajectory. An earlier per-prefix implementation was abandoned after the policy optimized rewarded prefixes instead of full completion. This shaping was used only for training; evaluation used each benchmark's native scoring.

\paragraph{Stability adaptations.} Two rollout-time issues shaped the recipe. (i) \emph{Context blow-ups}: a few tools return very large payloads (full spreadsheets, paper bodies), so we offload large tool outputs to files the agent can \texttt{grep}, trading turns for context. (ii) \emph{Looping}: the model sometimes repeated malformed tool calls (e.g., passing a string where a server demanded a number) until it exhausted context; we compact corrected retry sequences and retain only the most recent reasoning trace in the rolling context, matching the model's chat template and reducing context bloat.

\paragraph{Hyperparameters.} SFT: max length 131{,}072, AdamW (\texttt{betas}=$0.9/0.95$), LR $1\!\times\!10^{-5}$ cosine, warmup ratio $0.05$, weight decay $0.01$, 1 epoch. RL: max sequence length 80K, GSPO advantage estimator, 8 samples per prompt, rollout temperature $1.0$, LR $5\!\times\!10^{-5}$ constant, KL coefficient $0$ (KL-loss coefficient $0.001$, low-variance $k3$ estimator), entropy coefficient $0.001$, gradient clip $1.0$, 8 actor nodes $\times$ 8 H200 GPUs. Infrastructure was built on \texttt{slime}, Megatron-LM \citep{megatronlm2019}, and SGLang \citep{sglang2024}; all evaluations served the model in BF16 with SGLang. Additional configuration details are in Appendix~\ref{app:hparams}.

\section{Evaluation Protocol}
\label{sec:protocol}

\paragraph{What counts as evidence.} The holdout (40 in-family tasks) is an in-distribution \emph{pipeline check}: it confirms that the pipeline changes target-environment performance, but it shares the environment's degrees of freedom and cannot by itself establish transfer. We report five external benchmarks: Toolathlon \citep{toolathlon2025}, \tnb \citep{tau2bench2025}, BFCL-V4 \citep{bfcl2025}, SWE-Bench Pro \citep{swebenchpro2025}, and Terminal-Bench~2 \citep{terminalbench2026}. No external-benchmark task or grader appears in SFT or RL. After the initial base-family and teacher choices, no external score informed the training reward, hyperparameters, trained-checkpoint selection, or early stopping.

\paragraph{Overlap and contamination.} The training collection was Toolathlon-inspired and reused some MCP server implementations and tool schemas from its repository. Toolathlon leaderboard performance was considered when choosing the initial base-model family and SFT teacher, but no Toolathlon task instance or grader entered training, and later benchmark results did not determine the reward, training hyperparameters, trained checkpoint, or stopping. The other benchmarks also share capabilities and, in places, tool/API surfaces with \env. We cannot audit the base model's pretraining corpus, so we do not claim the benchmarks were unseen during pretraining. Our empirical claim is narrower: under matched evaluation conditions, the trained checkpoint differs from the base after post-training that excluded the external benchmark tasks and graders.

\paragraph{Sampling and statistics.} The primary results use greedy decoding and each benchmark's native scoring and aggregation. Base and trained checkpoints were evaluated with the same protocol within each benchmark. We report descriptive point estimates: the benchmarks are correlated, several use native weighted aggregation, and a single post-training run does not estimate training-run variance, so we do not attach a formal cross-benchmark hypothesis test. We also retain scores reported as \passk{4} in the accepted-paper snapshot, obtained from four temperature-$1.0$ trajectories and the archived harness's native aggregation. The raw units needed to identify these values with a standard unweighted any-of-four estimator are unavailable, so we refer to them only as native four-sample diagnostics. The retained \tnb configuration aggregates Airline, Retail, Telecom, and Banking. These diagnostics are distinct from the updated greedy evaluation.

\section{Results}
\label{sec:results}

\paragraph{Transfer to external tool-use benchmarks.} Table~\ref{tab:main} reports the base and trained models on the holdout and three external tool-use evaluations. The updated greedy evaluation gives a $+17.5$\pp\ holdout change, $+9.6$\pp\ on Toolathlon, and $+3.5$\pp\ on BFCL-V4. We additionally retain the accepted-paper \tnb result ($+5.3$\pp) and four-sample diagnostics, while separating their provenance from the updated greedy scores.

\begin{table}[t]
\centering
\small
\caption{Tool-use results (\%); $\Delta$ is base$\to$trained. Greedy values are from the updated final evaluation, except \tnb, which is retained from the accepted-paper snapshot. Four-sample diagnostics are from that earlier snapshot and should not be treated as decompositions of the updated greedy scores.}
\label{tab:main}
\begin{tabular}{lccc}
\toprule
 & Base & Trained & $\Delta$ \\
\midrule
\multicolumn{4}{l}{\emph{\passk{1} (greedy)}}\\
LHMTA holdout           & 10.0 & 27.5 & $+17.5$ \\
Toolathlon              & 22.2 & 31.8 & $+9.6$ \\
\tnb                    & 54.8 & 60.1 & $+5.3$ \\
BFCL-V4                 & 55.7 & 59.2 & $+3.5$ \\
\midrule
\multicolumn{4}{l}{\emph{Accepted-paper four-sample diagnostic}}\\
Holdout                 & 15.9 & 37.2 & $+21.3$ \\
Toolathlon              & 25.9 & 36.1 & $+10.2$ \\
\tnb                    & 57.1 & 62.9 & $+5.8$ \\
BFCL-V4                 & 67.7 & 72.2 & $+4.5$ \\
\bottomrule
\end{tabular}
\end{table}

\paragraph{Transfer to software-engineering benchmarks.} The strongest cross-domain test is improvement on software work absent from \env. The same trained checkpoint improves on SWE-Bench Pro by $+5.8$\pp\ and Terminal-Bench~2 by $+2.8$\pp\ (Table~\ref{tab:coding}). The claim concerns the direction of change from general-agent post-training, not absolute state of the art. Across the five reported external evaluations, all aggregate changes are positive; we treat that consistency descriptively rather than as a formal test because the benchmarks are correlated and \tnb comes from the retained evaluation snapshot.

\begin{table}[t]
\centering
\small
\caption{Updated cross-domain software-engineering results (greedy \passk{1}, \%). The training collection contains no software-engineering tasks; $\Delta$ is base$\to$trained.}
\label{tab:coding}
\begin{tabular}{lccc}
\toprule
 & Base & Trained & $\Delta$ \\
\midrule
SWE-Bench Pro       & 20.5 & 26.3 & $+5.8$ \\
Terminal-Bench~2    & 44.9 & 47.7 & $+2.8$ \\
\bottomrule
\end{tabular}
\end{table}

\paragraph{Interpreting the point estimates.} The results are descriptive, not uncertainty-free. Greedy decoding removes sampling temperature from the primary comparison, but finite task sets, native benchmark weighting, evaluation infrastructure, and training-run variation remain. We ran one full post-training experiment at this scale and do not have the raw paired units needed for benchmark-appropriate confidence intervals across all rows. We therefore emphasize the direction and magnitude of the observed changes and avoid claims of statistical significance.

\paragraph{Per-category structure and a capability-vs-attempt caveat.} Toolathlon gains in the accepted-paper diagnostic snapshot are uneven across categories (Table~\ref{tab:toolathlon-cat}). The largest is Office, $0.0\to16.5$\pp. This zero baseline leaves greater room for changes in task initiation and completion, behaviors that were also encouraged by the low-score effort floor; $16.5\%$ absolute remains low. Finance ($+10.1$), Shopping ($+11.5$), and Tech ($+9.9$) improve from non-trivial baselines, which is less readily explained by task initiation alone.

\begin{table}[t]
\centering
\small
\caption{Toolathlon \passk{1} by category (\%) from the accepted-paper diagnostic snapshot; these values do not arithmetically decompose the updated aggregate in Table~\ref{tab:main}.}
\label{tab:toolathlon-cat}
\begin{tabular}{lccccccc}
\toprule
 & Academia & Campus & Daily & Finance & Office & Shopping & Tech \\
\midrule
Base    & 30.7 & 15.4 & 21.4 & 25.0 & 0.0  & 50.0 & 26.7 \\
Trained & 38.1 & 19.5 & 29.2 & 35.1 & 16.5 & 61.5 & 36.6 \\
$\Delta$ & $+7.4$ & $+4.1$ & $+7.8$ & $+10.1$ & $+16.5$ & $+11.5$ & $+9.9$ \\
\bottomrule
\end{tabular}
\end{table}

\paragraph{BFCL slice structure.} In the retained \passk{1} breakdown, BFCL-V4 improves on Agentic ($41.8\to46.9$, $+5.1$\pp), Multi-Turn ($57.1\to62.1$, $+5.0$\pp), and Live ($73.1\to75.7$, $+2.6$\pp), while Non-Live and Hallucination regress slightly. We report these as diagnostics without uncertainty estimates.

\section{Behavioral Analysis: Recurring Differences}
\label{sec:behavior}

Aggregate scores tell us that measured performance changed; they do not tell us how behavior differed. We compared base and trained trajectories for the same task under the same prompt, environment, scaffold, and evaluation protocol, focusing on fail-to-pass pairs across \env, Toolathlon, SWE-Bench Pro, and Terminal-Bench~2. An automated root-cause-analysis agent (Claude Opus 4.8) traced failures to evidence-supported divergences; the authors then checked candidate reports against the raw reasoning, tool calls, environment responses, edits, tests, and artifacts. The four categories below were developed iteratively as an interpretive framework, not as mutually exclusive labels or claims about internal model mechanisms.

We organize the observed differences with a simple goal-directed loop: set a goal $\to$ take an action $\to$ update working state from feedback $\to$ compare to the goal $\to$ repeat, applied recursively as high-level goals decompose into sub-goals and concrete tool calls. Tool use, normal completion, and tool diversity were directly shaped for low-scoring trajectories by the effort floor. Precise local-goal formation, faithful state construction, parent-goal stability, and test-based verification were not explicit reward terms.

\paragraph{1. Forming the correct goal at every scale.} The trained model grounds local goals more carefully in the actual environment state. On a Toolathlon clickstream A/B-test task, the base model defined conversion as \texttt{clicks/store\_views} (yielding rates above 100\%), while the trained model related the columns to the event order and used \texttt{store\_views/clicks}. In a SWE-Bench Pro NodeBB task, both models found helpers and tests encoding the required selected-field semantics. The base reimplemented filtering in backend paths and changed missing-object behavior; the trained model formed the narrower goal of routing the public methods through the established helpers. The surfaces differ, but both cases concern translating a parent objective and observed state into the correct local target.

\paragraph{2. Building goal-relevant working state.} On a Toolathlon task computing year-over-year growth from a workbook, the base model treated formula strings as missing data and discarded values needed by the report; the trained model reopened the workbook using cached formula values. In a SWE-Bench Pro Ansible task, the base introduced a local helper that called \texttt{keyword.iskeyword} even though only \texttt{iskeyword} was imported, then changed the test setup after the resulting \texttt{NameError}. The trained model connected the requirement to the repository's central collection-name validator and reused that path. In both cases, the difference is whether observed environment structure enters the working representation that guides the next action.

\paragraph{3. Keeping parent goals stable through local repairs.} On a course-scheduling holdout task, the base model first recognized the ``only 2xxx-level courses'' constraint, then violated it (selecting a 1xxx course) while fixing a credit-cap problem, and then falsely self-certified. In a SWE-Bench Pro PowerShell CLIXML task requiring decoded control characters to be preserved, the base model encountered a stale test expecting the old behavior and ``fixed'' it by reintroducing exactly the behavior it was told to change; the trained model recognized the test as stale relative to the new requirement. The pressure, holding the parent requirement fixed while repairing a local problem, is domain-independent.

\paragraph{4. Verifying completion at each boundary.} On a library-catalogue holdout task, the base model generated a partial 44-book intermediate file, read it back, and declared the catalogue complete (checking an artifact against itself); the trained model reconciled two workbook views and confirmed both contained the same 347 authors before proceeding. In a SWE-Bench Pro refactor that moved logic out of a class component, the base model checked only that the new helpers were exported, while the trained model searched for stale call sites and ran a targeted type-check over the changed files. The behavior is verifying the property the downstream task actually depends on, not a convenient proxy.

\paragraph{Quantifying the behaviors.} Table~\ref{tab:behavior} reports deterministic metrics over 731 matched SWE-Bench Pro task pairs. With similar retrieval volume, the trained model repeats less retrieved information. Its changed files overlap the reference patch more often and its patches add far fewer lines, patterns consistent with more targeted investigation and editing but not proof of correctness. Verification changes most clearly: the share of runs containing a formal test rises from $37.5\%$ to $73.3\%$.

\begin{table}[t]
\centering
\small
\caption{Behavioral metrics over 731 matched SWE-Bench Pro task pairs. Reference patches are not unique ground truth, and smaller patches are not inherently better.}
\label{tab:behavior}
\begin{tabular}{@{}p{2.15in}ccp{1.55in}@{}}
\toprule
Metric & Base & Trained & Interpretation \\
\midrule
Mean retrieval calls & 25.6 & 23.9 & similar retrieval effort \\
Repeated retrieved information & 22.5\% & 14.3\% & less repeated investigation \\
Reference-patch files touched & 2.69 & 3.08 & greater reference overlap \\
Mean added lines & 415.5 & 111.7 & smaller edit footprint \\
Runs containing a formal test & 37.5\% & 73.3\% & more verification \\
\bottomrule
\end{tabular}
\end{table}

Retrieval calls are read-only inspection commands. Repeated information is the share of lowercased, overlapping 12-token spans already seen in an earlier retrieval output from the same run. Reference-patch overlap and added lines are computed from the final submitted diff, and formal tests are commands invoking standard test runners.

\paragraph{Capability or propensity?} One reading is that post-training mainly teaches the model to \emph{attempt} tasks and emit gradeable artifacts. The effort floor explicitly encouraged tool use, normal completion, and tool diversity for low-scoring trajectories, so changes in attempting and finishing may partly reflect direct shaping; we did not run a no-floor ablation. Precise local-goal formation, faithful state construction, parent-goal stability, and test-based verification were not explicit reward terms. Their recurrence across domains and the aggregate correlates in Table~\ref{tab:behavior} are consistent with broader transfer, but do not establish a causal mechanism.

\section{Discussion: Designing for Transferable Behavior}
\label{sec:discussion}

Reading transfer behaviorally has a direct payoff for data design. If the unit of value is a \emph{task pressure} (a way the environment forces the agent to form a goal, build state, hold requirements, or verify), then surface diversity (different prompts, tools, file types) can be a trap: a dataset can look diverse while clustering on one or two failure modes, and two superficially unrelated tasks can train the same transferable pressure. This mirrors the instruction-tuning shift from ``more data'' to ``data selected for the reasoning it exercises'' \citep{activeit2023,deita2024,less2024}, now at the level of environments rather than prompt-response pairs. A practical next step is a flywheel: build tasks that produce meaningful failures, tag the pressure each is meant to exercise, train, evaluate on held-out, nearby, and far benchmarks, analyze trajectories rather than only score deltas, and use the observed pattern to generate testable hypotheses for the next data round. One such hypothesis is that adding tasks requiring preservation of mature-system contracts may improve code transfer; the present experiment does not test it.

\section{Conclusion}

The value of an agentic post-training environment is not only how high it pushes its own score but whether the ways of working it encourages are reusable. We post-trained an open MoE model on long-horizon, MCP-based professional workflows and observed positive changes across five external evaluations, including two software-engineering benchmarks despite no software-engineering training tasks. We then provided an exploratory behavioral account: more careful local-goal formation, faithful state-building, goal stability under repair, and evidence-based verification recur across office workflows and code. Explicit effort shaping can account for some changes in attempting and finishing; the remaining behavioral observations and aggregate correlates are consistent with, but do not prove, a broader mechanism of transfer. Together, cross-benchmark evaluation and trajectory-level analysis offer a practical way to study not only what agents achieve, but how their behavior changes.

\bibliography{references}
\bibliographystyle{colm2026_conference}

\appendix
\vspace{-\baselineskip}
\section{Reported Training Configuration}
\label{app:hparams}
SFT (framework: \texttt{ms-swift}): max length 131{,}072; optimizer AdamW (fused), betas $(0.9,0.95)$; LR $1\!\times\!10^{-5}$, cosine schedule, warmup ratio $0.05$; weight decay $0.01$; max grad norm $1.0$; 1 epoch; batch size 4; gradient accumulation 4. SFT corpus: 3{,}000 Kimi-K2.6 trajectories, rejection-sampled at reward $>0.9$.

RL (GSPO): max sequence length 80K; RoPE scaling factor 8; context-parallel size 4; expert-parallel size 8; tensor-parallel size 4; pipeline-parallel size 2; 8 actor nodes $\times$ 8 GPUs; SGLang TP/EP size 8; samples per prompt 8; rollout temperature $1.0$, top-$p$ $1.0$; LoRA targets \texttt{q,k,v,o,gate,up,down}, rank 32, $\alpha$ 32; LR $5\!\times\!10^{-5}$ constant; Adam betas $(0.9,0.98)$, eps $1\!\times\!10^{-8}$; weight decay $0.1$; clip $1.0$; KL coefficient $0$; KL-loss coefficient $0.001$ (low-variance $k3$); entropy coefficient $0.001$; rollout routing replay on; BF16. SFT trajectory generation and RL used the same 363 training tasks; the 40 holdout tasks were excluded from both stages. Evaluation: BF16 on H200 with SGLang; primary \passk{1} greedy (temperature $0$); retained four-sample diagnostics at temperature $1.0$.

\vspace{-.5\baselineskip}
\enlargethispage{1.5\baselineskip}
\section{Continued Training and Task Saturation}
\label{app:continued}
Extending RL to a second and third epoch on the same tasks did not improve and eventually regressed: entropy fell from $\approx 0.9$ to $\approx 0.6$ by epoch two with declining training reward, and performance regressed further by epoch three. This is consistent with overfitting and suggests that additional task diversity may be more useful than additional passes over the same 363 tasks. On-policy distillation on self-generated successful trajectories \citep{onpolicydistill2024} is a natural next step.

\end{document}